\documentclass[twocolumn,amsmath,amssymb,showpacs,prl]{revtex4}
\usepackage{graphicx}
\usepackage{dcolumn}
\usepackage{bm}
\usepackage[english]{babel}
\usepackage{amssymb}
\usepackage{amsmath}
\usepackage{subfigure}
\usepackage{color}

\begin{document}

\title{
Microwave quantum optics as a direct probe of the Overhauser field
in a quantum-dot CQED device }

\author{Pei-Qing Jin$^1$}
\author{Jan Jeske$^2$}
\author{Andrew D. Greentree$^{2,3}$}
\author{Jared H. Cole$^2$}
\affiliation{$^1$Institute of Logistics Engineering, Shanghai Maritime University, Shanghai 201306, China}
\affiliation{$^2$Chemical and Quantum Physics, School of Applied Sciences, RMIT University, Melbourne 3001, Australia }
\affiliation{$^3$Australian Research Council Centre of Excellence for Nanoscale BioPhotonics}

\begin{abstract}
We show theoretically that a quantum-dot circuit quantum electrodynamics (CQED) device can be used as
a probe of the Overhauser field in quantum dots.
By coupling a transmission line to the interdot tunneling gate,
an electromagnetically-induced-transparency (EIT) scheme can be established,
whose Fano-type interference leads to
a sharp curvature in the reflection spectrum around resonance.
This sharp feature persists even in the presence of the fluctuating spin bath,
rendering a high-resolution method to extract the bath's statistical information.
For strong nuclear spin fields,
the reflection spectrum exhibits an Autler-Townes splitting,
where the peak locations indicate the strengths of the Overhauser field gradient (OFG).
\end{abstract}

\pacs{73.21.La, 78.67.Hc, 42.50.Gy}


\maketitle

\emph{Introduction}.
A singlet-triplet qubit (STQ) encoded in two-electron states
in a double quantum dot \cite{Levy},
has emerged as a promising candidate for
quantum-dot-based
quantum information processing systems \cite{Loss98,Burkard,Petta05,Foletti09,Barthel09,Studenikin}.
For quantum dots in III-V type semiconductors
the electrons interact with a large number of nuclear spins ($\sim 10^5$)
residing in the host lattice.
These nuclear spins provide a fluctuating magnetic field (Overhauser field)
with a statistical variance of a few mT \cite{Khaetskii,Merkulov02}.
This leads to a nanosecond-timescale dephasing time $T_2^*$ for the electron spins
\cite{Coish05,YaoW06}.
Both electrical and optical methods have been employed to mitigate this effects
\cite{Stepanenko,Bluhm,Gossard}.
Alternatively,
the spin bath can also assist in quantum information processing,
e.g., as a key ingredient in the universal control of a single STQ \cite{Foletti09},
or as a long-lived quantum memory \cite{Taylor03}.
Both of these aspects make it crucial to understand the Overhauser field
and the nuclear spin bath that generates it.

For the STQ in a double quantum dot,
the Overhauser fields are usually different between the two dots
due to, e.g., their geometric asymmetry.
This leads to an Overhauser field gradient (OFG).
So far, measurements of the OFG are based on a spin-to-charge conversion,
e.g., in Ref.~\cite{Petta05,Bluhm,Reilly08}.
The whole procedure involves several steps
and a cyclic sequence of gate-pulses is employed.
After initialization,
the system is moved to an operating point
where the STQ experiences a free precession under the OFG,
and then the system is brought back to its initial state
where its return probability reflects the OFG.
The statistical variance of the OFG is then extracted
from the Gaussian-like decay of the return probability.

Circuit quantum electrodynamics (CQED) devices
provide a bridge from solid state qubits to the photonic world.
By coupling qubits to a superconducting transmission line \cite{SupCQED},
fascinating quantum optic phenomena such as
single-qubit lasing \cite{SQL1,SQL2} and cooling \cite{SQC},
single photon generation \cite{Houck07} and detection \cite{Johnson10,Bozyigit},
have been achieved.
Our interest here is in electromagnetically induced transparency (EIT) \cite{EITRev},
which provides efficient control over optical responses of atomic systems
and thus gives rise to numerous applications
including precision measurement \cite{EITmea},
quantum optical memories \cite{EITmemory},
and coherent electron transfer \cite{Andrew04},
and was also demonstrated in CQED systems
\cite{EITinSC1,EITinSC2}.
Stimulated by success with CQED with superconducting qubits,
theoretical proposals towards quantum-dot based CQED devices
were put forward \cite{DR1,DR2,DR3,DR4,DR5}.
Recent fast experimental progress
includes atom-photon coupling in these hybrid systems \cite{DREXP1,DREXP2},
characterization of photon emissions \cite{Liu14}
and construction of a quantum dot maser \cite{Gullans}.

\begin{figure}[t]
\centering
\includegraphics[width=0.9\columnwidth]{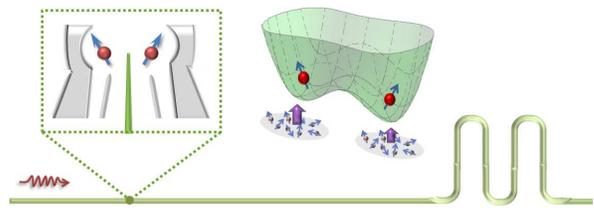}\\[4mm]
\caption{
Schematic of a quantum-dot CQED device.
A transmission line is coupled
to the interdot tunneling gate of a double quantum dot system
via a finger-shaped electrode extended
from the transmission line.
The quantum dots contain two electrons,
both of which experience effective magnetic fields
provided by the local nuclear spins.
 }\label{Fig:QDCQED}
\end{figure}
In this letter we propose an EIT-based method
to detect the nuclear spin field in a double quantum dot
via a CQED architecture, as shown in Fig.~\ref{Fig:QDCQED}.
The OFG breaks the spin conservation and mixes the STQ states.
When driving a probe field along the transmission line
coupled to the interdot tunneling gate,
the system can achieve EIT,
where due to a Fano-type interference
its reflection spectrum exhibits a double peak structure
with a sharp dip around resonance.
The fluctuating nuclear spin bath
imprints its statistical information
in the spectrum whose sharp feature provides
a high resolution detection mechanism.

\begin{figure}[t]
\centering
\includegraphics[width=0.93\columnwidth]{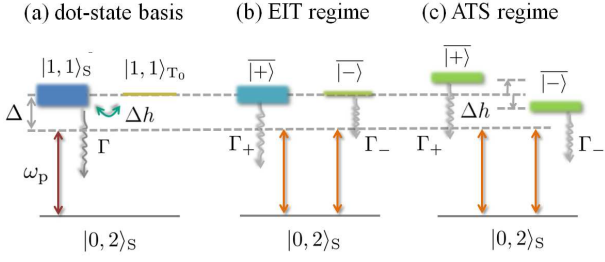}
\caption{
Panel (a):
Relevant energy levels in a dot-state basis.
The probe field with frequency $\omega_{\rm p}$
addresses the transition between the two singlets,
where a frequency detuning $\Delta$ is allowed.
An OFG with magnitude $\Delta h$ mixes two STQ states.
Charge noise renders a dominant decay channel
from state $|1,1\rangle_{\rm S}$
to $|0,2\rangle_{\rm S}$ with rate $\Gamma$.
Panel (b-c):
Dressed-state picture for an EIT and ATS regime, respectively.
The OFG couples the STQ states,
leading to two nuclear-spin-mediated states (NSMS),
$\overline{|\pm\rangle}$.
The NSMS have the same energy but distinctive widths
in the EIT regime (weak $\Delta h$),
while they develop different frequencies separated by $2\Delta h$
but the same width $\Gamma_\pm=\Gamma/2$
in the ATS regime (strong $\Delta h$).
 }\label{Fig:Energy}
\end{figure}

\emph{Model and approach}.
We consider a double quantum dot with two electrons.
The detection mechanism involves two STQ states,
$|1,1\rangle_{\rm S}$ and $|1,1\rangle_{\rm T_0}$,
in the $m_{\rm s} = 0$ subspace,
as well as a singlet with two electrons occupying
the right dot,
$|0,2\rangle_{\rm S}$.
We assume the system operates in a parameter regime
with negligible interdot tunneling
where the nuclear spins exert major influence on the STQ
\cite{Petta05,Bluhm}.
The two STQ states are then degenerate,
which by tuning gate voltages
can possess higher energy compared to the state $|0,2\rangle_{\rm S}$,
as shown in Fig.\ref{Fig:Energy} (a).

The energy difference between the two singlet states,
$|1,1\rangle_{\rm S}$, and $|0,2\rangle_{\rm S} \}$,
denoted as $\omega_0$
(throughout the paper we set $\hbar =1$),
depends on the dot detuning and on-site Coulomb interaction \cite{Burkard}.
The STQ states are coupled via the OFG (with strength $\Delta h$).
Furthermore, we assume a transmission line
is coupled to the interdot tunneling gate \cite{DR5}.
A probe field with frequency $\omega_{\rm p}$
propagating through the transmission line then
addresses the two singlets,
$|1,1\rangle_{\rm S}$ and $|0,2\rangle_{\rm S}$,
with Rabi frequency $\Omega_{\rm p}$.
Then the Hamiltonian in the space spanned by
$\{ |0,2\rangle_{\rm S}, |1,1\rangle_{\rm S}, |1,1\rangle_{\rm T_0} \}$,
is given by
\begin{eqnarray}\label{Eq:H0}
H_0 &=&
\left(
\begin{array}{ccc}
 0  & \Omega_{\rm p}  \cos(\omega_{\rm p}t)  & 0   \\[2mm]
 \Omega_{\rm p}  \cos(\omega_{\rm p}t)  & \omega_0  & \Delta h        \\[2mm]
 0   & \Delta h   & \omega_0
  \end{array}
\right).
\end{eqnarray}
This Hamiltonian indicates that,
as compared to a conventional situation \cite{EITRev},
the quantum-dot CQED system
actually constitutes a $\Lambda$ configuration
where an EIT scheme can be established.
Here the OFG effectively plays the role of a dc coupling field,
which is resonant with the corresponding electronic transition.

We analyze the dynamics of the system in a master-equation formalism
with dissipation described by Lindblad operators \cite{DRLasing},
\begin{eqnarray}\label{Eq:ME}
 \dot{\varrho} = -i[H_{\rm RF}, \varrho]  + \frac{\Gamma}{2}
 \left( 2\,\sigma_-\,\varrho\,\sigma_+ -\varrho\,\sigma_+ \sigma_- -\sigma_+ \sigma_-\, \varrho
 \right),
\end{eqnarray}
where $H_{\rm RF}$ denotes an effective Hamiltonian in a rotating frame
transformed under a unitary matrix
$U_{\rm RF} = \exp\left\{(-it){\rm diag}(0, \Delta, \Delta)\right\}$
with detuning $\Delta = \omega_0 -\omega_{\rm p}$,
as well as a usual rotating wave approximation \cite{Carmichael},
$\varrho$ represents the reduced density matrix for the dot system,
and $\sigma_-\equiv |0,2\rangle_{\rm S} {}_{\rm S}\langle1,1|$.
We focus on a dominant relaxation process
from $|1,1\rangle_{\rm S}$ to $|0,2\rangle_{\rm S}$ caused by charge noise.
The pure dephasing of the state $|1,1\rangle_{\rm S}$
can be included whose rate merely adds to the relaxation rate \cite{EITRev}.
For the double quantum dot,
its dephasing rate is of a few hundred MHz \cite{Wallraff14}.
Note that
the relaxation from $|1,1\rangle_{\rm S}$
to $|1,1\rangle_{\rm T_0}$ is strongly suppressed
due to spin conservation.
This helps to protect the resulting EIT window
which would otherwise be deteriorated by such additional dissipation.

\begin{figure}[t]
\centering
\includegraphics[width=0.85\columnwidth]{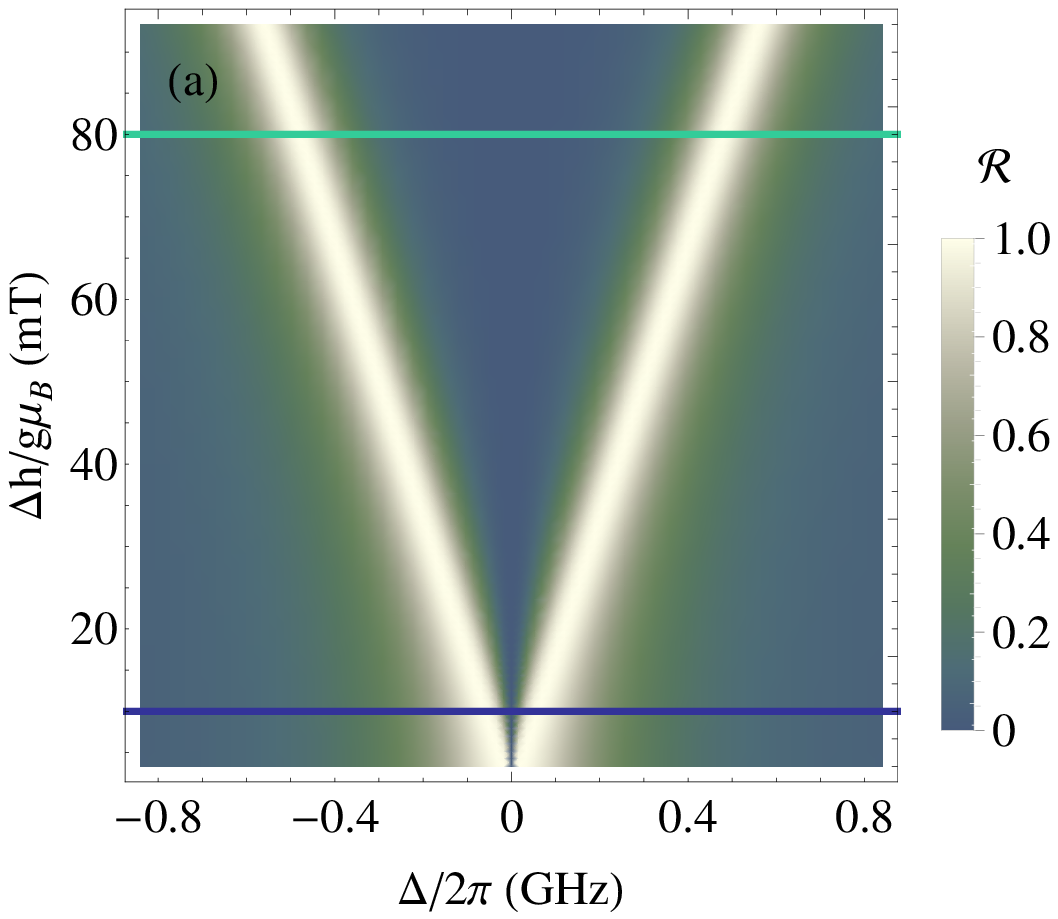}\\[4mm]
\includegraphics[width=0.82\columnwidth]{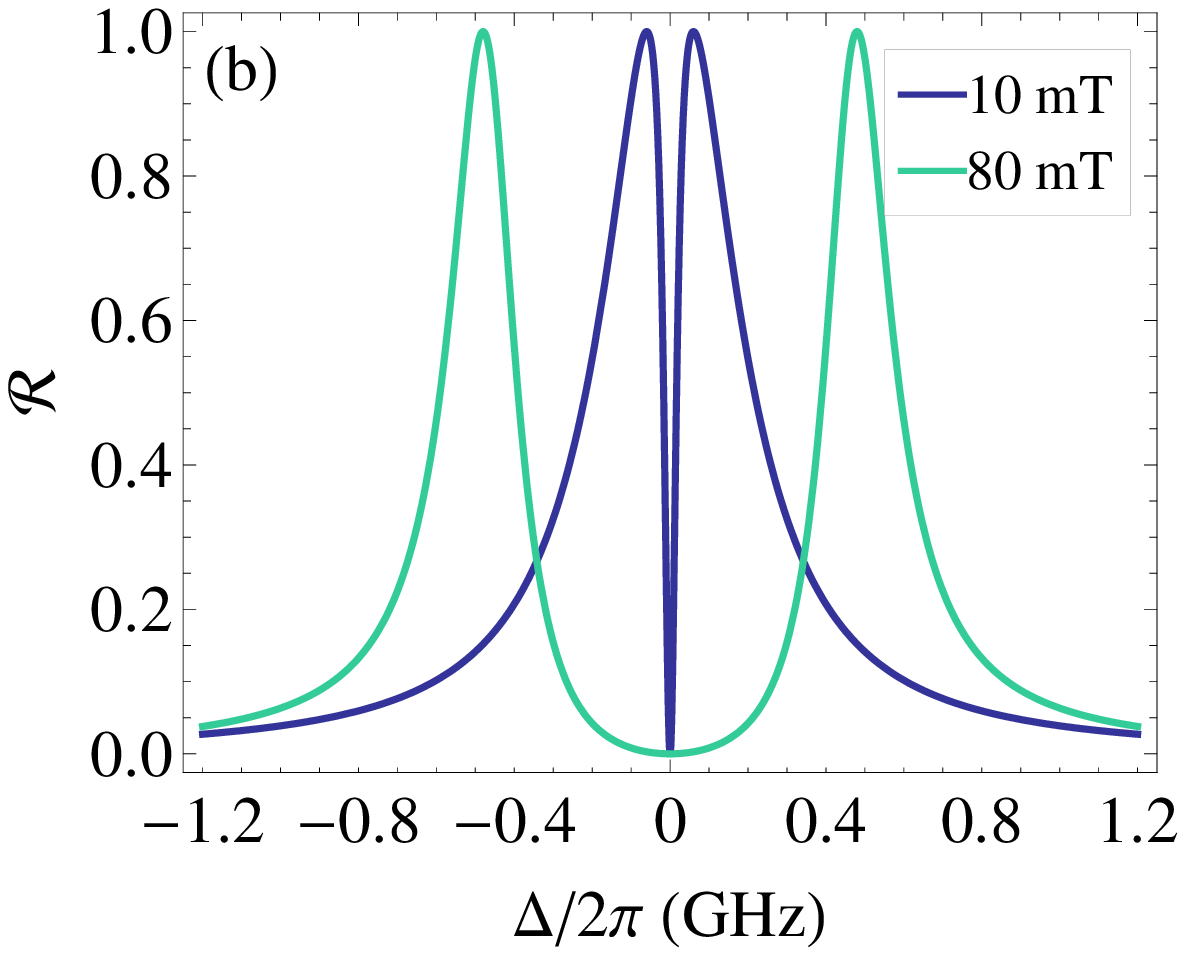}
\caption{
Panel (a):
Density plot of reflection $\mathcal R$
as a function of both detuning and OFG.
The spectrum exhibits a double-peak structure.
As $\Delta h$ increases,
the spectral peaks move far apart,
and the system develops from an EIT regime
to an ATS regime.
Panel (b):
Reflection as a function of detuning.
For weak OFG with $\Delta h/(g\mu_{\rm B}) = 10 {\rm mT} $ (blue line),
the spectrum exhibits a sharp dip around the resonance $\Delta =0$
due to a Fano-type interference.
For strong OFG with $\Delta h/(g\mu_{\rm B}) = 80 {\rm mT} $ (green line),
the peaks are well separated,
and the spectrum becomes rather flat at the resonance.
 }\label{Fig:AS}
\end{figure}

\emph{EIT Scheme}.
Equation~(\ref{Eq:H0}) suggests from the level-structure perspective
that an EIT scheme can be established in the coupled system.
Another prerequisite condition
relies on the relative strength between the OFG and the electronic decay rate.
It essentially determines whether the system is in an EIT or
Autler-Townes-splitting (ATS) regime \cite{Anisimov}.
For illustration we adopt a dressed-state picture
which was developed in the study of the Fano-type interference
in an EIT scheme \cite{Cohen92,EITRev}.
In this picture, the EIT phenomenon results from the destructive interference
between two scattering processes through two dressed states.
In our case the OFG plays the role of a dc coupling field,
we refer to these dressed states as ``nuclear-spin-mediated'' states (NSMS).
These states, $\overline{|\pm\rangle}$,
are eigenstates of a non-Hermitian effective Hamiltonian
including the electronic decay rate
in the subspace of $\{|1,1\rangle_{\rm S},|1,1\rangle_{\rm T_0}\}$,
whose eigenvalues are
$\Delta_\pm = -(\pm\sqrt{16\,\Delta h^2-\Gamma^2}+i\Gamma)/4$.
For weak OFG ($\Delta h/\Gamma \ll 1$),
the NSMS are approximately those STQ states modified by small hybridization,
namely, $\overline{|\pm\rangle} \simeq |1,1\rangle_{\rm S/T_0} \pm \alpha |1,1\rangle_{\rm T_0/S} $
with $\alpha = -2\Delta h/(i\Gamma)$.
In this case, as shown in Fig.~\ref{Fig:Energy} (b),
the NSMS have the same frequency ${\rm Re}[\Delta_\pm] \simeq 0$,
but with distinct widths:
one is close to the natural width of $|1,1\rangle_{\rm S}$,
namely, $\Gamma_+ = -{\rm Im}[\Delta_+] \simeq \Gamma/2 - 2\Delta h^2/\Gamma $;
the other is much narrower,
$\Gamma_- = -{\rm Im}[\Delta_-] \simeq 2\Delta h^2/\Gamma $.
Then the scattering processes through either of these NSMS interfere destructively with each other,
leading to a cancellation in the spectrum at resonance.
This interference was identified as a Fano-type interference \cite{Cohen92}.
For strong OFG compared to the decay rate,
the NSMS are energetically shifted by $\pm\Delta h$,
each with width $\Gamma/2$, as shown in Fig.~\ref{Fig:Energy} (c).
The scattering processes can then be viewed as
through two separate channels, resulting in two Lorentzian line shapes.
This spectral structure is then an ATS,
which in our case is a nuclear-spin-field-induced Stark effect.

\emph{Reflection Spectrum}.
We use the reflection spectrum as a monitor of the OFG.
Following a similar procedure as in Ref.\cite{EITinSC1},
we find the reflection coefficient
$r = r_0 (-i\Gamma/\Omega_{\rm p})
{}_{\rm S}\langle 1,1|\varrho|0,2\rangle_{\rm S}$,
where $r_0$ denotes a bare reflection coefficient with vanishing OFG.
We then introduce $\mathcal R = |r/r_0|^2$
to characterize a renormalized strength of the reflection coefficient.
By seeking the steady-state solution of Eq.~(\ref{Eq:ME}),
we obtain
\begin{eqnarray}\label{Eq:refl}
\mathcal R = \frac{\Delta^2(\Gamma/2)^2}
 {(\Delta^2-\Delta h^2)^2+\Delta^2(\Gamma/2)^2}.
\end{eqnarray}
Figures ~\ref{Fig:AS} (a-b) show the reflection spectrum
as a function of detuning for different values of OFG.
There are a few features of the spectrum
which will be employed in our detection strategy.
First, the spectrum exhibits a double-peak structure
with maxima at $\Delta_{\rm max} = \pm\Delta h$.
Hence the strength of the OFG
can be extracted from the peak positions.
Second,
a sharp dip appears at the resonance $\Delta =0$ for small OFG
[blue line in Fig.~\ref{Fig:AS} (b)].
It is due to the Fano-type interference in the EIT regime.
We approximate its detuning-dependence as
$|\partial {\mathcal R }/\partial \Delta |
 \simeq [\Gamma^2/(2~\Delta h^4)]\Delta $,
which clearly indicates an increasing sensitivity
for weak OFG.
This sharp curvature can then serve as a promising basis
for a high-resolution detection scheme.
Besides,
as the OFG grows,
the spectral peaks move further apart
as shown in Fig.~\ref{Fig:AS} (a).
For strong OFG,
either peak develops into a Lorentzian line shape
with width $\Gamma/2$
[green line in Fig.~\ref{Fig:AS} (b)].
In this case the system is in an ATS regime,
where the spectrum becomes rather flat around resonance.

Fluctuations in the OFG modify the reflection spectrum.
So far we have treated the nuclear spin field
as a static field,
which due to its slow dynamics
is valid for a single run of experiment.
During the data collection,
however, the Overhauser field fluctuates,
leading to various magnitudes of the OFG for each independent experiment,
the distribution of which takes a Gaussian form,
$ P(\Delta h) = (\sqrt{2\pi} \sigma)^{-1}
 \exp\left[{-(\Delta h-h_0)^2/(2\sigma^2)}\right]$ \cite{Merkulov02,Coish05}.
Its mean value $h_0$ depends on the polarization of the nuclear spins,
which is typically a few mT for an unpolarized nuclear spin bath \cite{Petta05},
and can reach hundreds of mT via external polarization mechanism \cite{Bluhm}.
The statistical variance $\sigma$ can be of several tenths of $\mu{\rm eV}$.
The final result of the reflection
is then based on an ensemble average
over different nuclear spin configurations,
given by,
\begin{eqnarray}\label{Eq:EnsAve}
\Bigl<\mathcal{R}\Bigr>_{\rm nucl}
 &=& \frac{\sqrt{\pi}~\Delta\Gamma} {4\sqrt{2}~\sigma}
 {\rm Re}\biggl\{Z_{\rm p}^{-1}
  \Bigl[e^{-X_-^2}\Bigl(1+{\rm Erf}(iX_-)\Bigr)
        \nonumber \\[2mm]
 &&
 +e^{-X_+^2}\Bigl(1-{\rm Erf}(iX_+)\Bigr) \Bigr] \biggr\},
\end{eqnarray}
where ${\rm Erf(x)}$ is the error function,
$Z_{\rm p} = \sqrt{\Delta} \sqrt{\Delta+i(\Gamma/2)}$
and $X_\pm = (\mp Z_{\rm p}-h_0)/(\sqrt{2}~\sigma)$.
As shown in Figures~\ref{Fig:AS} (c-d),
the variance $\sigma$ has a broadening effect on the spectral peaks.
As $\sigma$ grows,
contributions to the averaged reflection spectrum
come from a wider range of OFG.
In Ref.\cite{Andew09} the effect of inhomogeneous broadening from frequency detuning
on EIT phenomenon was studied.
In contrast here the broadening effect arises from
the fluctuating strength of the coupling field, namely, the OFG.

\begin{figure}[t]
\centering
\includegraphics[width=0.9\columnwidth]{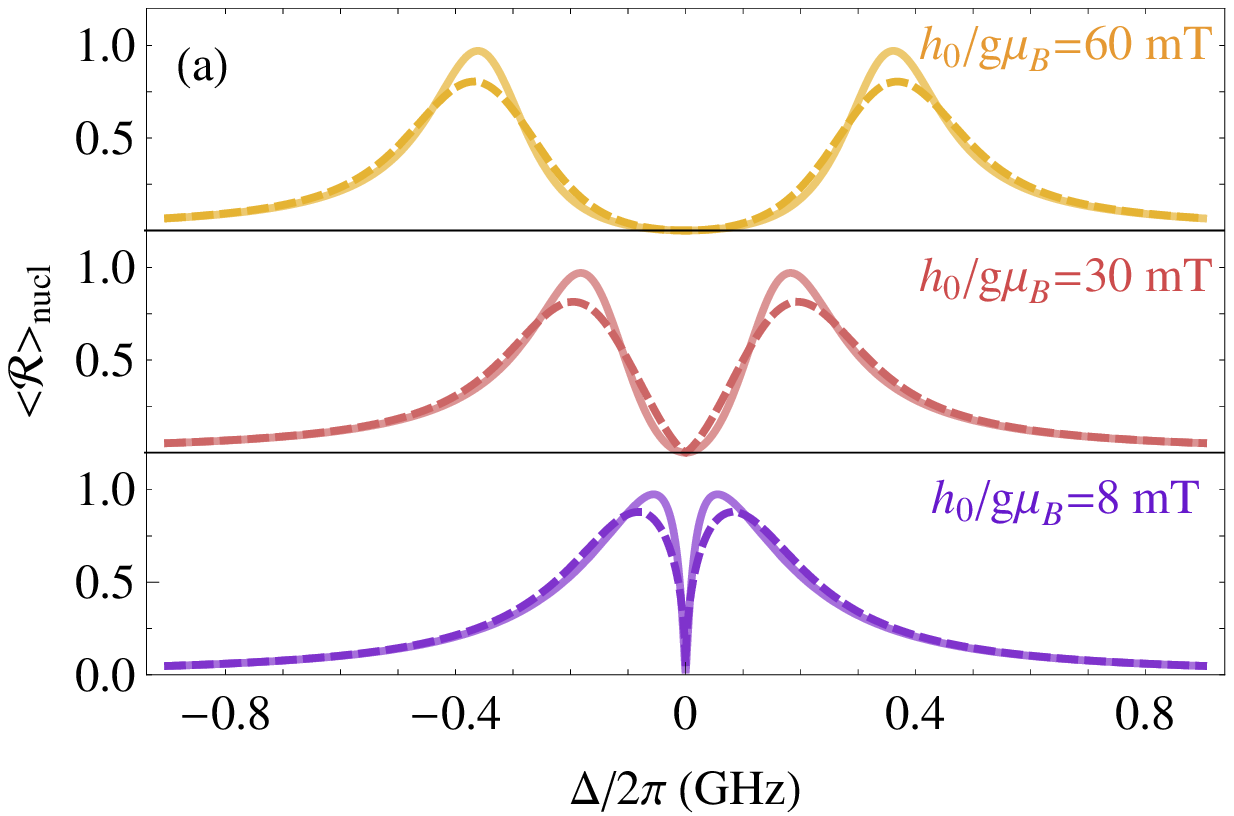}\\[4mm]
\includegraphics[width=0.9\columnwidth]{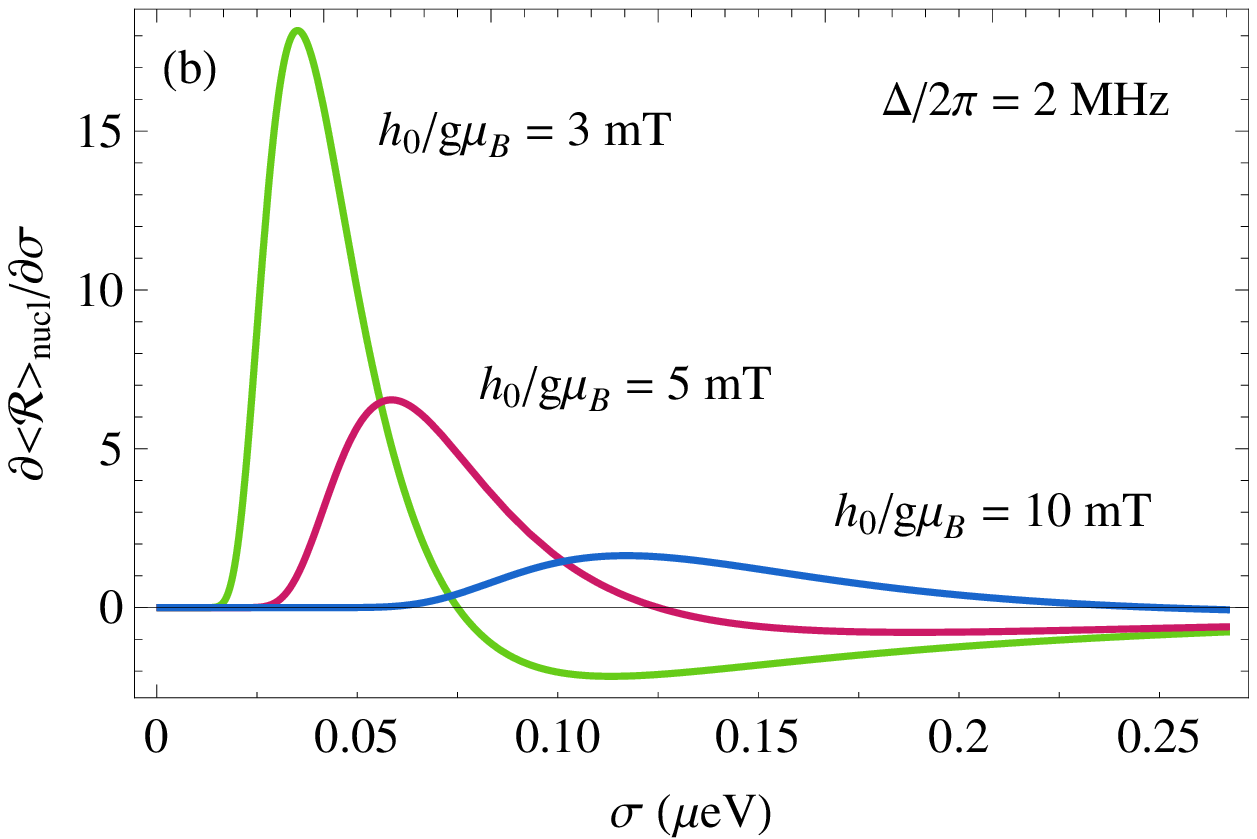}
\caption{
Panel (a):
Ensemble-averaged reflection $<\mathcal R>_{\rm nucl}$
as a function of detuning with OFG variance $\sigma = 75 \rm{neV}$ (solid lines).
and $\sigma = 250 \rm{neV}$ (dashed lines).
The spectral peaks get broadened by increasing variance.
For small mean value $h_0/(g\mu_{\rm B}) = 10 {\rm mT}$
the sharp feature around the resonance persists even
in the presence of variance.
As the mean value increases,
the spectrum starts to develop a flat profile at the resonance.
Throughout the paper we choose $\Gamma/2\pi= 400 {\rm MHz}$,
and consider GaAs quantum dots with $|g|=0.44$.
Panel (b):
Reflection sensitivity on statistical variance $\sigma$
for different mean values $h_0$ at detuning $\Delta/2\pi = 2 {\rm MHz}$.
This sensitivity reduces as increasing the mean value,
where the spectrum becomes rather flat around the resonance.
 }\label{Fig:average}
\end{figure}

\emph{Detection mechanism}.
A striking feature of the ensemble-averaged reflection
is that the sharp curvature around the resonance in the EIT regime
still persists in the presence of the variance,
as shown in Fig.~\ref{Fig:AS} (c).
We then adopt it as a sensitive measure to extract the variance $\sigma$.
We expand $\Bigl<\mathcal{R}\Bigr>_{\rm nucl}$ around the resonance,
which can be approximated as,
$ \Bigl< \mathcal{R} \Bigr>_{\rm nucl}^{(\rm L)}
= \sqrt{\pi\Gamma|\Delta|}/(2\sqrt{2}\sigma) \exp[-h_0^2/(2\sigma^2)]$.
Its sensitivity on the variance is
\begin{eqnarray}
 \frac{\partial \Bigl< \mathcal{R} \Bigr>_{\rm nucl}^{(\rm L)} }{\partial \sigma}
 =  \left(\frac{\sqrt{\pi\Gamma|\Delta|}}{2\sqrt{2}} \right)
 \left( \frac{h_0^2 - \sigma^2}{\sigma^4} \right) \exp\left(-\frac{h_0^2}{2\sigma^2}\right).
\end{eqnarray}
As shown in Fig.~\ref{Fig:AS} (d),
the sensitivity reaches its maximum around $\sigma \simeq h_0/2$.
For small variance $\sigma \ll h_0$,
the broadening effect of the fluctuating OFG is weak,
and the spectrum slope close to the resonance
is still very sensitive to the variance,
rendering a high detection resolution.
However, as the mean value $h_0$ grows,
the sensitivity reduces.
This is because
the system in this parameter regime
falls into the ATS scheme,
where the double peaks are well separated
and the spectrum is rather flat around resonance.
In this situation,
the above detection mechanism is not available any more.
Instead, one can return to the peak structure.
Around each peak,
the spectrum can be approximated as a Lorentzian profile.
For weak variance, its ensemble-averaged result
still locates around $\pm h_0$,
with peak value $1-4~\sigma^2$.
The strategy is then to extract the variance from the peak value.
%
%

Except for the statistical variance,
the mean value of the OFG, $h_0$,
can also be extracted from the spectrum,
namely, from the peak locations,
$ h_0 \simeq \Delta_{\rm max}$.
This approximation works quite well for small variance,
where its deviation,
$|(h_0-\Delta_{\rm max})/h_0|$, is below ten percent.
As the variance grows,
the peak shoulders become broadened,
and the peak locations shift to higher frequencies,
which deteriorates the resolution.

\emph{Discussion and summary}.
We showed an EIT scheme can be established
in a quantum-dot CQED device
by coupling a transmission line
to the interdot tunneling gate.
The sharp spectral curvature in the EIT window
allows for a high-resolution mechanism
to extract the statistical information of the nuclear spin bath.
According to the order of the OFG strength,
different detection strategies are required for an optimal resolution.
For STQ experiments without external polarization of nuclear spins \cite{Petta05,Coish05},
the OFG is usually several mT in GaAs quantum dots,
much weaker compared to its charge-noise induced decay rate of
several hundreds of MHz.
The system is then in the EIT regime
where the sharp spectral curvature is available for the detection.
When the nuclear spins are polarized with OFG to the order of hundreds of mT \cite{Foletti09,Bluhm},
the system is in the ATS regime,
and the bath information is better resolved from the peak structures instead.

The ensemble-averaged result in Eq.~(\ref{Eq:EnsAve})
is based on the Gaussian distribution of the OFG,
where we assumed the two nuclear spin baths are uncorrelated
and the statistical variance $\sigma$ is of the same order as that of a single dot.
A partially correlated situation may arise
due to e.g., strong overlap between the two electron wavefunctions in each dot,
or dipolar contributions from spins in the crystal.
This would render a narrower statistical variance,
in which case our detection scheme becomes more sensitive.

We thank X.-M. Lu and P. Kotetes for helpful discussions.
PQJ acknowledges financial support from the National Natural Science Foundation of China (Grant No. 11304196),
as well as the Science and Technology Program of Shanghai Maritime University.
JJ and ADG acknowledge the support of the Australian Research Council (DP130104381).

\end{document}